\documentclass[preprint,12pt]{elsarticle}

\usepackage[utf8]{inputenc}
\usepackage{amssymb}
\usepackage{amsmath}
\usepackage{url}

\newcommand{\be}{\begin{eqnarray}}
\newcommand{\ee}{\end{eqnarray}}
\newcommand{\ba}{\begin{align*}}
\newcommand{\eea}{\end{align*}}
\newcommand{\hMpc}{{\ifmmode{h^{-1}{\rm Mpc}}
\else{$h^{-1}$Mpc}\fi}}

\journal{Astronomy and Computing}

\begin{document}

\begin{frontmatter}



\title{Field Realization Mass Function -- a code to investigate collapse barriers in non-Gaussian cosmologies}


\author{Sergey Pilipenko} 

\affiliation{organization={AstroSpace Centre of P.N. Lebedev Physical Institute of the Russian Academy of Sciences},
            addressline={Profsojuznaja 84/32}, 
            city={Moscow},
            postcode={117997}, 
            state={},
            country={Russia}}

\begin{abstract}
\texttt{FRMF} is a code which computes the dark matter halo mass function (MF) with several methods unified by the same principle: the MF is computed from a realization of a linear density field on a 3D grid by passing this density field through a barrier. This computation can be done in three ways: 1) exceedance fraction, which resembles the Press-Schechter method, 2) shoreline, which is an equivalent of excursion set methods, and 3) peak-patch, similar to the Bond and Myers 1996 proposition. For every method, a number of barriers can be applied: static, moving, and diffusing spherical collapse thresholds and ODE-based elliptical collapse criteria. The same density field can be used in \texttt{FRMF} and in simulations, enabling their direct comparison and reproducing the effects of box size, variance, and resolution in \texttt{FRMF}. The code can generate density fields for Gaussian and non-Gaussian cosmologies, including scale-dependent non-Gaussianity. Our analysis shows that the non-Gaussian case at high redshifts is a real challenge for most barriers. Across the tested methods, the best precision for the non-Gaussian MF can be obtained with the shoreline and peak-patch methods applied to the elliptical collapse barrier.
\end{abstract}




\end{frontmatter}



\section{Introduction}

The abundance of dark matter halos as a function of mass and redshift, $n(M,z)$, is one of the most direct links between the statistics of the primordial density field and observable large-scale structure. The straightforward way to obtain the MF is to run N-body simulations. However, simulations may be costly when one wants to reach high precision and a wide dynamical range and also to explore a large parameter space of cosmological models. This motivates the usage of analytical MFs and various simplified methods. These methods are usually built upon the theoretical idea of collapse: a patch of the Universe turns into a halo when it satisfies some criterion, which can be called a \textit{barrier}. Several approaches are summarized below, and the main motivation for this project was to find the ``right'' one, most suitable for scanning a parameter space of various cosmological models with scale-dependent primordial non-Gaussianity, especially focusing on high redshifts and small scales.

\subsection{Calculating the number density of halos}
\label{sec:criteria}
Three broad families of simulation-free methods are widely used to predict $n(M,z)$ from a linear density field:

\begin{itemize}
  \item \textbf{Exceedance fraction (Press--Schechter).} The original
  Press \& Schechter \cite{PS} construction identifies collapsed regions of mass $M$ with points of a field smoothed on a scale $R(M)$ that simply exceed a fixed threshold $\delta_R(\mathbf{x}) > \delta_c$. The halo mass function follows from   differentiating the volume fraction above the threshold with respect to mass, with the familiar factor-of-two cloud-in-cloud correction added by hand.

  \item \textbf{Excursion set (``shoreline'') methods.} \cite{BondColeEfstathiouKaiser1991}
  (BCEK) reinterpreted the Press--Schechter threshold-crossing as the first-passage
  problem of a random walk in the $(\delta,\sigma_R^2)$ space generated by smoothing the
  field on a decreasing sequence of scales, where $\sigma_R$ is the dispersion of the smoothed density field. This approach both derives the factor of two from
  first principles (for a sharp-$k$ filter) and resolves the cloud-in-cloud problem.
  In a field-level implementation this amounts to tracking, for every cell, the
  first scale at which the smoothed density crosses the barrier -- the ``shoreline''
  of the excursion set.

  \item \textbf{Peak-patch methods.} \cite{BondMyers1996} (BM96) instead assign
  halos to \emph{local density maxima} of the field rather than to all cells (or
  all excursion-set trajectories): candidate peaks are tested against a collapse
  criterion and grown outward until they are bounded by neighboring, already-claimed
  peaks, with each peak's mass set by the largest radius at which it still clears
  the barrier. This ensemble restriction to peaks, rather than to all positions,
  was also implemented in Excursion Set Peaks methods (ESP) \cite{ParanjapeSheth2012,ParanjapeShethDesjacques2013}.
\end{itemize}


\subsection{Barrier types}
\label{sec:barriers}
Independently of which of the three constructions above is used, the barrier
applied to the field can take several forms, of increasing physical detail:

\begin{itemize}
  \item \textbf{Static (spherical collapse).} The classic constant threshold
  $\delta_c \simeq 1.686/D(z)$, the linearly extrapolated overdensity for the
  collapse of a homogeneous sphere.

  \item \textbf{Moving barriers.} \cite{ShethMoTormen2001} (SMT) showed that the empirical Sheth--Tormen correction to Press--Schechter, which for a decade had
  been a pure fit to N-body mass functions, follows dynamically from combining the
  ellipsoidal-collapse condition of \cite{BondMyers1996} with the statistics of the
  most probable ellipticity of a Gaussian field at a given $\delta/\sigma_R$ \cite{Doroshkevich1970}. The result
  is a barrier that depends on the smoothing scale through $\sigma_R$

  \begin{equation}
  \label{eq:SMT}
      B(\sigma_R) =
  \sqrt{a}\,\delta_c\left[1+0.485(\sigma_R/\sqrt{a}\delta_c)^{1.23}\right],
  \end{equation} 
  where $a\approx0.707$ ($\sqrt{a}=0.84$) is calibrated from N-body simulations, while other constants are derived from elliptical collapse ODE solutions.

  \item \textbf{Diffusing barriers.} \cite{MaggioreRiotto2010b} proposed instead  that the Sheth--Tormen rescaling $\delta_c\to\sqrt{a}\,\delta_c$ is a statistical
  effect: the collapse threshold itself is not a fixed number, but diffuses around
  its mean (an Ornstein--Uhlenbeck process in $\sigma^2$), due to the unmodeled
  scatter from triaxiality and tidal fields, with the diffusion coefficient $D_B$
  calibrated against simulations ($a=1/(1+D_B)$). \cite{MaggioreRiotto2010c} extend
  this to non-Gaussian initial conditions (\S\ref{sec:barriers-png} below).

  \item \textbf{Elliptical (deformation) barriers.} Rather than folding the
  ellipsoidal-collapse condition into an effective, scale-dependent $B(\sigma)$, one
  can evaluate the exact ellipsoidal/ODE collapse criterion of \cite{BondMyers1996}
  directly on the eigenvalues of the (unnormalized) local tidal (deformation) tensor at each
  candidate peak. This was done in \cite{SandvikMollerLeeWhite2007} and in the excursion-set
  peaks barrier of \cite{CastorinaParanjapeHahnSheth2016} (ESP$\tau$), which adds the
  trace-free shear invariant $\tau$ to the isotropic ESP barrier of
  \cite{ParanjapeSheth2012,ParanjapeShethDesjacques2013}. 
\end{itemize}

\subsection{Scale-dependent primordial non-Gaussianity}
\label{sec:PNG}
The standard $\Lambda$CDM cosmological model assumes that the initial perturbations were Gaussian; however, many inflation theories predict Primordial Non-Gaussianity (PNG).
PNG is characterized at leading order by the bispectrum, the three-point function of
the primordial curvature (or Bardeen potential) perturbation, and is classified by
the shape this bispectrum takes as a function of the triangle formed by the three
wavevectors $k_1,k_2,k_3$ \cite{BabichCreminelliZaldarriaga2004,BartoloKomatsuMatarreseRiotto2004}.
\emph{Local} PNG peaks in the squeezed limit ($k_1\ll k_2\simeq k_3$) and arises
whenever the final curvature perturbation is a local, nonlinear function of one or
more Gaussian field(s) evaluated at the same point in space, as happens generically
in multi-field inflation, the curvaton scenario, and modulated-reheating models
\cite{Wands2010,BartoloKomatsuMatarreseRiotto2004}. \emph{Equilateral} and
\emph{orthogonal} PNG instead peak for triangles with $k_1\simeq k_2\simeq k_3$ 
\cite{BartoloKomatsuMatarreseRiotto2004,BabichCreminelliZaldarriaga2004}. A
\emph{folded} (flattened) shape, peaking for $k_1\simeq k_2\simeq k_3/2$, arises in
models with a non-Bunch--Davies initial vacuum state
\cite{BabichCreminelliZaldarriaga2004}.

Among these, local-type PNG has historically received the most attention and is by
far the most relevant to the simulation and halo-statistics literature that follows.
Physically, this is because it is the generic outcome of essentially any
multi-field mechanism, not a feature special to a narrow class of models
\cite{Wands2010}, and because its squeezed-limit enhancement has a uniquely clean
observational signature: it induces a scale-dependent correction to the large-scale
halo and galaxy bias, $\Delta b_1(k)\propto f_\mathrm{NL}(b_1-1)/k^2$, which diverges
on the largest scales and has become the single most constraining large-scale-structure
probe of PNG \cite{DalalDoreHutererShirokov2008}. The local PNG is
conventionally parametrized, in the Bardeen-potential convention adopted throughout
this paper, by a single dimensionless amplitude $f_\mathrm{NL}$ through the quadratic
ansatz
$$\Phi(\mathbf{x}) = \varphi(\mathbf{x}) + f_\mathrm{NL}\left[\varphi(\mathbf{x})^2 - \langle\varphi^2\rangle\right],$$
where $\varphi$ is an auxiliary Gaussian primordial potential field
\cite{BartoloKomatsuMatarreseRiotto2004,PlanckCollaboration2020PNG}. The current
tightest bound on this amplitude, from the Planck CMB bispectrum, is
$f_\mathrm{NL}^\mathrm{loc}=-0.9\pm5.1$ ($68\%$ CL) \cite{PlanckCollaboration2020PNG},
fully consistent with the single-field slow-roll expectation $f_\mathrm{NL}\sim\mathcal{O}(n_s-1)$,
but still leaving ample room for the scale-dependent amplitudes
discussed below.

Most of the PNG literature summarized above is phrased for a constant $f_\mathrm{NL}$.
However, many inflationary mechanisms predict a non-Gaussianity amplitude that runs
with scale, $f_\mathrm{NL}(k)$, rather than a single number. DBI and other
non-canonical single-field models with a time-varying sound speed generate a
power-law running of the equilateral amplitude,
$f_\mathrm{NL}^{\rm eff}(k)\propto k^{-2\kappa}$ \cite{LoVerdeMillerShanderaVerde2008},
with an analogous general power-law ansatz for arbitrary shapes parametrized by a
running index $n_\mathrm{NG}$ \cite{SefusattiEtal2009}. A more strongly
scale-dependent possibility arises when a spectator field's mass transitions from
heavy to light during inflation, suppressing the bispectrum on CMB scales while
leaving it unsuppressed at the much smaller scales relevant to galaxy and
cluster-mass halos \cite{RiottoSloth2011}; phenomenological versions of this idea
(power-law running, constant-with-cutoff, Gaussian bump, and tanh-saturated
templates) have been used to fit the excess abundance of bright galaxies seen by
JWST at $z\gtrsim8$--$10$ \cite{HabouzitEtal2014,BiagettiFranciolniRiotto2022}, and
negative scale-dependent $f_\mathrm{NL}$ at $k\gtrsim0.1\,h\,\mathrm{Mpc}^{-1}$ has
been proposed as a route to suppressing small-scale power and easing the $S_8$
tension \cite{StahlEtal2024b,BaldiEtal2024}. These scale-dependent templates evade
the CMB bispectrum bound essentially by construction (the amplitude is small or
zero at CMB pivot scales) and are only constrained directly by LSS- and
simulation-scale probes.  The absence of $\mu$ distortion in FIRAS data allows for $f_{NL}<3.6\times10^6$ at small scales \cite{BianchiniFabbian2022}.

\subsection{Consistency of the collapse barrier and the non-Gaussian bias}
\label{sec:consistency}
Besides the global mass function itself, the important prediction of non-Gaussian models is the additional scale-dependent bias, $b_\phi$ defined as (see, e.g., \cite{Sullivan25}):
\begin{equation}
    \delta_h = b_1\delta + f_{NL}b_\phi \varphi,
\end{equation}
where $\delta_h$ is the density fluctuation of halos, $b_1$ is the usual bias (Gaussian, Lagrangian) and $\varphi$ is the primordial potential defined in \S\ref{sec:PNG}. In \cite{DesjacquesJeongSchmidt2017} it was shown that the consistency relation $b_\phi = \partial\ln\bar{n}_h/\partial\ln\sigma_8$,
guaranteed on general grounds by the peak-background split, holds for a Lagrangian halo model if and only if the collapse barrier depends only on physical (un-normalized) fields, with no explicit dependence on $\sigma_8$ or $\sigma_R$.
Moving barriers such as the Sheth--Tormen form, the diffusing barrier of
\cite{MaggioreRiotto2010b}, and other stochastic-barrier models of this kind all have explicit dependence on $\sigma$ and therefore violate the relation, especially at small masses (below the typical one). Barriers evaluated directly on physical tidal
eigenvalues, such as the \cite{SandvikMollerLeeWhite2007} barrier and the ESP$\tau$
barrier of \cite{CastorinaParanjapeHahnSheth2016}, satisfy it. 

This matters
specifically for the PNG studies we are interested in: a barrier that gives an accurate
$n(M,z)$ by construction (through scale-dependent fitting parameters) can still give
the wrong non-Gaussian bias response if it is not built from physical fields.

\subsection{Previous attempts to apply these barriers to non-Gaussian initial conditions}
\label{sec:barriers-png}

\cite{MatarreseVerdeJimenez2000} (hereafter MVJ) obtained the first analytic non-Gaussian mass function via a path-integral / saddle-point argument applied to the exceedance fraction, introducing an effective threshold $\delta_\ast$ and a fudge factor absorbing uncaptured physics. 
This was followed by a calculation based on an Edgeworth expansion of the density PDF \cite{LoVerdeMillerShanderaVerde2008}, hereafter the LV formula. \cite{Grossi} derived an all-orders integral modification of the MVJ formula, hereafter called the Grossi formula. 

The authors of \cite{Grossi} also compared analytical results with simulations for $|f_{NL}|\leq 1000$ in the mass range $M>10^{12}$ M$_\odot/h$ and found that the MVJ formula works better for large negative $f_{NL}$ while their own variant is better for large positive values. Analysis of scale-dependent PNG models may be interesting for even larger values of $f_{NL}$, since FIRAS data on the CMB spectrum give $f_{NL}<3.6\times10^6$ at small scales \cite{BianchiniFabbian2022}. Baldi et al. 2024 \cite{BaldiEtal2024} compared simulations with MVJ predictions for small-scale $f_{NL}$ up to $10^4$ and found that it requires adjusting the fudge factor depending on $f_{NL}$ and redshift, which limits using the MVJ approximation as a prediction tool when there are no simulations available.

On the excursion-set side, \cite{MaggioreRiotto2010c} showed that the naive non-Gaussian generalization of
the Press--Schechter factor of two is not simply justified. 
\cite{LamSheth2009} combined a moving
barrier (Sheth--Tormen-like) with non-Gaussian initial conditions and found a further
small correction from the barrier shape itself, concluding that the assumed barrier
height matters quantitatively once $|f_\mathrm{NL}|\gtrsim100$. In \cite{DAmicoMussoNorenaParanjape2011} the excursion set approach was used to derive a more precise formula than MVJ and LV which also includes explicit theoretical error bars. The authors used a constant barrier of modified height, with a correction motivated by the diffusing barrier approach. Finally, \cite{MussoSheth2014} demonstrated a Monte-Carlo realization of non-Gaussian (Lognormal) random walk trajectories.

Moving barriers are based on the Doroshkevich formula \cite{Doroshkevich1970} which describes the distribution of shear deformation tensor eigenvalues for the Gaussian case. In \cite{Lam09} it was shown that for weak non-Gaussianity only the trace of the tensor, i.e. the overdensity $\delta$, is affected by non-Gaussianity, while the ellipticity/prolateness distribution remains untouched for a fixed $\delta$ to linear and quadratic order in $f_{NL}$. This justifies the usage of a moving barrier for PNG models, but limits it to small $f_{NL}$.

The illustration that a barrier calibrated on Gaussian simulations does not
simply carry over to the non-Gaussian case comes from the diffusing-barrier
framework. \cite{AchitouvCorasaniti2012} extended the \cite{MaggioreRiotto2010b}
diffusing barrier to a Diffusive Drifting Barrier (DDB) with both a diffusion
coefficient $D_B$ and a mean drift $\beta$, calibrated on Gaussian N-body halos. When
the same model is confronted with non-Gaussian simulations, both $D_B$ and $\beta$
are found to be themselves $f_\mathrm{NL}$-dependent: in their N-body calibration,
$D_B$ rises from $\simeq0.33$ at $f_\mathrm{NL}=0$ to $\simeq0.63$ at
$f_\mathrm{NL}=800$, and the Gaussian-calibrated parameters only remain adequate over
a limited range, roughly $-50\lesssim f_\mathrm{NL}\lesssim150$. In other words, the
width of the barrier's scatter -- not just its mean -- must be refit against
non-Gaussian simulations for the diffusing-barrier approach to remain accurate,
which is also the physical origin of the $f_\mathrm{NL}$-dependent fudge factors $q$
seen in the LoVerde-type and Matarrese--Verde--Jimenez-type formulas. 

\subsection{Motivation}
As summarized above, analytical approximations require calibration against simulations for scale-dependent PNG with $|f_{NL}|>10^3$, thus they lack predictive power in the regime of interest for some scale-dependent $f_{NL}$ models. Especially, this is important at high redshifts: as $z$ grows, the more and more rare peaks form halos, so their abundance is more sensitive both to the non-Gaussian effects, and the approach used to count halos. This motivated me to search for more complicated methods, than the existing analytical formulas. Existing excursion set calculations for PNG models I am aware of also involve analytical approximations, not a direct Monte-Carlo calculation of trajectories, except for the \cite{MussoSheth2014} paper. The latter, however, uses a same-scale, pointwise transform of the PDF (Gaussian to Lognormal), not a full computation of trajectories with bispectrum-induced correlations following from the quadratic ansatz for PNG I'm interested in.

To solve this problem, I have chosen the straightforward approach: to generate realizations of the density field on a 3D grid and to apply various existing collapse criteria to it. This approach gives a side effect: unlike excursion set simulations, field realizations can be compared to simulations directly, i.e. the same field is used for a simulation and for an approximate MF calculation. These considerations motivated the creation of \texttt{FRMF}. 

A similar idea, to use a field realization in order to compute the MF, was implemented in the \texttt{PINOCCHIO} code \cite{monaco2002pinocchio,monaco2013pinocchio,munari2017pinocchio}, however \texttt{FRMF} differs substantially from it. In \texttt{PINOCCHIO}, the barrier is a variant of the elliptical family, the collapse condition is computed from Lagrangian perturbation theory (LPT) instead of solving ODEs like BM96. Then for each grid cell the earliest collapse time is computed, and the collapsed grid cells are treated as particles and grouped together using their Eulerian positions calculated from LPT. The \texttt{PINOCCHIO} approach allows one to trace individual halos, their merger histories and compare them with simulations. \texttt{FRMF} is simpler, it calculates only the MF (however, the peak-patch method also outputs individual halo masses).

The arguments of \cite{DesjacquesJeongSchmidt2017} on the bias consistency relation point out that some of the barriers widely used in the literature, such as the SMT one, may not be suitable for the PNG MF calculation at all by construction. This proposition also deserves testing, and one of the motivations for \texttt{FRMF} was to compare different barriers.

The paper is organized as follows. The brief description of the \texttt{FRMF} methods and barriers, and the construction of PNG initial conditions are given in \S\ref{sec:methods}. The main results are presented in \S\ref{sec:comparison}: it describes a simulation suite that was used for the comparison with the code. In \S\ref{sec:sim-mf} and \S\ref{sec:sim-bias} the mass function and the non-Gaussian bias parameter predicted by various methods implemented in \texttt{FRMF} are compared with simulations. In \S\ref{sec:sim-fc} an attempt is made to further improve the agreement with simulations by adjusting the barrier height factor $f_c$. The final quantitative comparison of \texttt{FRMF} with simulations and with analytical MF approximations is presented in \S\ref{sec:sim-approx}. The conclusions are given in \S\ref{sec:conclusion}.

\section{Methods}
\label{sec:methods}
The \texttt{FRMF} code is published on github: \url{https://github.com/spilipenko/FRMF}. It consists of several modules aimed at the generation of initial density fields, application of barriers introduced in \S\ref{sec:barriers} to the three methods presented in \S\ref{sec:criteria}, computation of analytical approximations, and also utility functions. The code itself is written in Julia, it utilizes thread-based parallelization as well as GPU acceleration, if available. The detailed description of all the algorithms, step by step, is given in the code documentation, here I give only a brief description of what \texttt{FRMF} actually does.

\subsection{Initial conditions}
\label{sec:ics}
Gaussian initial conditions (ICs) are created using the method of Pen 1997 \cite{Pen1997}, i.e. by convolving the random white noise field with the density kernel computed from the cosmological power spectrum. ICs for the local PNG models are generated using the standard approach following, e.g., \cite{DalalDoreHutererShirokov2008,monofonic}. Specifically, the non-Gaussian potential field is generated as
\begin{equation}
    \phi_\mathrm{NG}(\mathbf{x}) = \phi_G(\mathbf{x}) + \phi_\mathrm{corr}(\mathbf{x}) - \langle\phi_\mathrm{corr}\rangle,
\end{equation}
where the correction is computed in Fourier space as:
\begin{equation}
    \hat\phi_\mathrm{corr}(\mathbf{k}) = f_\mathrm{NL}(k) \cdot \widehat{\phi_G^2}(\mathbf{k})
\end{equation}
and $\widehat{\phi_G^2}(\mathbf{k}) = \mathrm{FFT}[\phi_G^2(\mathbf{x})](\mathbf{k})$ is the Fourier transform of the squared field. In real space, $\phi_\mathrm{corr}(\mathbf{x}) = \mathrm{IFFT}[\hat\phi_\mathrm{corr}]$.

Squaring $\phi_G$ in real space produces a field $\phi_G^2$ whose Fourier modes extend out to $2k_\mathrm{Nyq}$. On the bare $N$ grid those excess modes fold back (alias) onto the resolved modes, contaminating the non-Gaussian correction. To get rid of this aliasing, zero-padding is used as proposed in \cite{monofonic,Stahl22,Adame26}.
In addition to the standard single ICs, the code supports generation of paired ICs as proposed in \cite{Angulo16} to reduce realization skewness. 

The resulting ICs can be saved in a format which can then be used by the \texttt{ginnungagap} code to generate ICs for simulations in Gadget-2 or other formats.

\subsection{Mass function calculation}
\label{sec:mf}
Given the initial density field, the code computes the mass function (MF) by three methods introduced in \ref{sec:criteria}. The first and the simplest is the exceedance fraction. In this method, a logarithmic grid in sphere radius $R$ is created (from $3L/N$ to $L/4$ by default), for each radius the density field is smoothed with the real-space top-hat filter of radius $R$ using a Fourier transform, the number of cells exceeding the threshold is computed, multiplied by 2 and differentiated with respect to mass to obtain the mass function for the mass grid corresponding to the grid in $R$. When the constant barrier is used, this method reproduces the Press-Schechter mass function.

In the second, shoreline method, a new 3D grid is added: the shoreline mask. The loop over the grid of filter radii is traversed from larger to smaller $R$, and at each step any cell passing the barrier (i.e. exceeding the threshold) is marked. The fraction of marked cells is considered as the fraction of matter in halos at mass $>M(R)$, and the mass function is constructed by differentiating this  fraction.

In the third method, called peak-patch, in addition to the mask grid, one more 3D grid is introduced, called $g_{prev}$, and also a 1D array of peak candidate radii $R_{max}$, whose role is discussed below. The main calculation is done again in a loop of decreasing filter radii $R$. The goal, according to the BM96 \cite{BondMyers1996} proposition, is to find peaks and for each peak find the maximal radius $R_{max}$ at which the peak satisfies the barrier criterion. Neighboring peaks may appear close to each other, and they are cleaned on the next step: starting from peaks with the largest $R_{max}$ every peak ``claims'' a sphere of radius $R_{max}$ around it. Already claimed cells cannot be claimed by other peaks. The peaks that cannot claim a certain fraction of their initial volume inside $R_{max}$ are discarded.

In the \texttt{FRMF} implementation of this method at each step in filter radius $R$ all density maxima are found by comparing each grid cell to its 26 neighbors. Then for each cell at a maximum the barrier is evaluated and the threshold margin $g=\delta-\delta_c$ is found. If $g>0$, the peak's maximum radius is found by interpolating $R$ between two adjacent steps in $R$, using $g$ stored in $g_{prev}$ at the previous step in $R$. After finding all peaks passing the barrier,  the claiming procedure starts: peaks are sorted in descending order of $R_{max}$, each peak checks how many not yet marked cells inside $R_{max}$ around it are available. If the fraction of the volume of available cells inside the $R_{max}$ sphere is below 10\% (this fraction is a parameter), no cells are marked and the peak is discarded. Otherwise, cells are marked and their number defines the final halo mass. At the end of the loop over $R$, we have a list of halo masses and the MF is computed by binning them.

The \texttt{FRMF} implementation of the peak-patch method differs from the BM96 one: in the latter the peaks were identified for a single smoothing radius, and then for each peak $R_{max}$ is found by growing it and smoothing the field in real space. Using the grid of $R$ improves the code performance. Also BM96 used a grid with only $128^3$ cells and were interested in cluster-mass halos. \texttt{FRMF} covers a larger dynamical range, so there is no justification for using a single smoothing radius to identify the peaks.

The $R$ grid in \texttt{FRMF} is generated covering the range from 3 cells ($3L/N$) to 1/4 of the box size ($L/4$). This corresponds to the mass range from $M_{min}$ to $M_{max}$. However, a special resolution investigation has shown that for the peak-patch method the MF gets lower by up to 60\% at $M_{min}$, and the safe range starts at $3 M_{min}$ where the deviation reaches 2\% at $z=5$ and 10\% at $z=0$. This was found by comparing two runs with $N=512$ and $N=1024$.

The implementation of the first two barriers from \S\ref{sec:barriers} is straightforward: the static barrier just compares the cell density with the constant $\delta_c$, while the moving barrier compares with the SMT formula (\ref{eq:SMT}).
The diffusing barrier of \cite{MaggioreRiotto2010b} is slightly more complicated: the barrier is computed individually for each cell on a 3D grid. First, the barrier is initialized with $\delta_c$ in each cell. The barrier value  is updated at each step in decreasing smoothing radius $R$:
\begin{equation}
    B_\mathbf{x}^{(i+1)} = B_\mathbf{x}^{(i)} + \sqrt{2D_B ( \sigma_R^{(i)}-\sigma_R^{(i+1)})}\;\mathcal{N}(0,1),
\end{equation}
where $\mathcal{N}(0,1)$ is a normally distributed random number. This barrier has one free parameter, the diffusion coefficient $D_B$. In \cite{MaggioreRiotto2010b} the analytic derivation gives $D_B=0.25$, while the calibration against the SMT MF gives $D_B=0.41$. We used the latter value as the default.

The next family of elliptical barriers includes three methods which require the calculation of the deformation tensor:
\begin{equation}
    \Psi_{ij}(\mathbf{x};R) = \frac{2a}{3\Omega_{m,0}H_0^2}\frac{\partial^2 \Phi_R}{\partial x_i \partial x_j},
\end{equation}
where $a$ is the scale factor, and $\Phi_R$ is the potential smoothed with the top-hat filter. The eigenvalues of the tensor are $\lambda_1 \leq \lambda_2 \leq \lambda_3$, and its trace is the overdensity: $\lambda_1 + \lambda_2 + \lambda_3 = \delta_R(\mathbf{x})$.

The first variant of an elliptical barrier was extracted from Sandvik et al. \cite{SandvikMollerLeeWhite2007}:
\begin{equation}
\label{eq:sandvik}
    \delta_\mathrm{ell}(\lambda_1, \lambda_2, \lambda_3) = \delta_c\!\left[1 + \alpha_1(\lambda_2 - \lambda_1)^{\alpha_2} + \alpha_3(2\lambda_3 - \lambda_2 - \lambda_1)^{\alpha_4}\right]
\end{equation}
with calibrated parameters $\alpha_1 \simeq 0.2809$, $\alpha_2 = 1.3557$, $\alpha_3 \simeq 0.070$, $\alpha_4 \simeq 1.41205$. Equation (\ref{eq:sandvik}) is a fit to the solutions of the ellipsoidal collapse equations introduced in BM96 \cite{BondMyers1996}. The same equations were used in the SMT paper \cite{ShethMoTormen2001}, and they also provide a fit $\delta_\mathrm{ell}$ as a function of ellipticity $e_v=(\lambda_3-\lambda_1)/(2\delta_R)$ and prolateness $p_v=(\lambda_3+\lambda_1-2\lambda_2)/(2\delta_R)$. I've tried to use this SMT fit instead of the Sandvik one and found a very minor difference.

As an alternative, \texttt{FRMF} can also solve the elliptical collapse equations of BM96 by itself (using a solver from \cite{rackauckas2017differentialequations}) and provide a fit which is tested in the region of eigenvalues not covered by the Sandvik or SMT fits. Specifically, the collapse criterion is parametrized as a function of $v=\lambda_2-\lambda_1$ and $w=2\lambda_3-\lambda_2-\lambda_1$ and the collapse equations are solved on a grid $v/\delta_c\in[0,4]$, $w/\delta_c\in[0,8]$. The resulting fit has a maximum deviation from the ODE solutions of 1.6\% over this range, while the Sandvik et al. formula reaches 16\%.

A simplified version of the barrier which takes into account the ellipsoidal collapse was proposed in \cite{CastorinaParanjapeHahnSheth2016} and called ESP$\tau$ in \texttt{FRMF}. This barrier utilizes the traceless part of the deformation tensor, the barrier details are given in the \texttt{FRMF} documentation.  

As noted in \cite{ShethMoTormen2001}, elliptical collapse gives a barrier which is always higher than the spherical-collapse barrier, $\delta_{ell}\geq \delta_c$, which results in the total suppression of halo abundance. To avoid this and to fit the simulated MFs, SMT introduced a barrier height factor of 0.84 -- they multiply the barrier height by this value. To allow this modification, all elliptical-type barriers in \texttt{FRMF} accept a parameter $f_c$ set to 1 by default. As we will show below, in the exceedance and shoreline methods one has to set $f_c=0.84$ to get a reasonable agreement with simulations, while the peak-patch method does not require this correction. On the other hand, the equations of elliptical collapse used in BM96 \cite{BondMyers1996} formally give a solution which has a barrier height of $0.95\delta_c$ in the spherical limit. BM96 treat this as a defect of the method and re-calibrate the results to $\delta_c$ in this limit. We will discuss this 0.95 factor later in \S\ref{sec:sim-fc}.

\section{Comparison with simulations}
\label{sec:comparison}
The comparison of the \texttt{FRMF} results with $N$-body simulations allows us to assess the accuracy of the various mass-function calculation methods presented in the previous sections. For this purpose I prepared a suite of 48 simulations. The simulation parameters were chosen to match those of \cite{BaldiEtal2024} and \cite{StahlEtal2024b}. The suite covers three scale-dependent PNG models together with their Gaussian reference realizations. All simulations contain $512^3$ dark-matter particles and share the same input $\Lambda$CDM power spectrum and cosmological parameters, $\Omega_\mathrm{m}=0.3175$, $\Omega_\Lambda=0.6825$, $\Omega_\mathrm{b}=0.049$, $h=0.6711$ and $n_\mathrm{s}=0.9624$. Snapshots are saved at $z=5$, $3$, $1.5$ and $0$. The white-noise fields are generated by the \texttt{FRMF} code, and the initial particle displacements are produced with \texttt{ginnungagap} \cite{ginnungagap} using the Zel'dovich approximation at $z_\mathrm{init}=99$. Thirty-two of the simulations are used for the mass-function comparison and have $\sigma_8=0.834$; the remaining 16 vary $\sigma_8$ by $\pm 5\%$ (i.e.\ $\sigma_8=0.793$ and $0.877$) to measure the response of the halo mass function to the normalization of the power spectrum, i.e.\ the PNG bias parameter $b_\phi$. 

Three scale-dependent $f_\mathrm{NL}(k)$ shapes are used: 
\begin{enumerate}
\item ``Baldi-type'', a tanh-squared model $f_\mathrm{NL}(k)=f_\mathrm{NL}^{\rm max}\,\tanh^2(\alpha k/k_0)$ with $f_\mathrm{NL}^{\rm max}=10^4$, normalized to $f_\mathrm{NL}(k_0)= 10$ at $k_0=0.05\,\mathrm{Mpc}^{-1}$ \cite{BaldiEtal2024};

\item same shape, but with negative $f_\mathrm{NL}^{\rm max}=-10^4$ (normalized to $f_\mathrm{NL}(k_0)=-10$ at $k_0=0.05\,\mathrm{Mpc}^{-1}$);

\item ``Stahl-type'', a smooth step $f_\mathrm{NL}(k)=-300\,\bigl[1+\tanh\bigl((k-0.15)/0.1\bigr)\bigr]$ (with $k$ in $h\,\mathrm{Mpc}^{-1}$), which goes from $0$ on large scales to $-600$ on small scales \cite{StahlEtal2024b}.
\end{enumerate}
The parameters in which the simulations differ are listed in Table~\ref{tab:sims}.

\begin{table}[t]
\centering
\caption{Simulation parameters. \label{tab:sims}}
\begin{tabular}{lcc}
\hline
 & Baldi-type & Stahl-type \\
\hline
Box size $L$ [$h^{-1}\,\mathrm{Mpc}$] & 100 & 500 \\
Particle mass [$h^{-1}\,M_\odot$] & $6.57\times10^8$ & $8.21\times10^{10}$ \\
PNG model & tanh, $f_\mathrm{NL}=\pm10^4$ & step, $f_\mathrm{NL}=-600$ \\
Realizations per variant & 8 & 4 \\
Mass-function runs & 24 & 8 \\
$\sigma_8$-ladder runs ($b_\phi$) & 16 & 0 \\
\hline
\end{tabular}
\end{table}

The simulations were run with the code \texttt{BanyanPM} \cite{banyanpm}, a GPU Tree-PM code with behavior identical to \texttt{GADGET-2} \cite{springel}. \texttt{BanyanPM} was tested on $512^3$-particle simulations with both Gaussian and PNG initial conditions in exactly the same setup as used in this \texttt{FRMF} comparison. The halos were extracted using the \texttt{Rockstar} halo finder \cite{Behroozi} and spherical-overdensity masses of isolated halos are used.

The comparison is done in two parts: the MF comparison of simulations against various methods, and the bias parameter $b_\phi$ comparison. Each of the three MF calculation methods implemented in \texttt{FRMF} was run with the 6 available barriers. The barriers were run at their default parameters, except for three elliptical barriers: Sandvik (deformation), ESP$\tau$ and ODETable. As noted in Section~\ref{sec:mf}, there is a motivation to lower the barrier height by $f_c=0.84$ in the case of the exceedance and shoreline methods for the deformation barriers, which we do. All the \texttt{FRMF} calculations use the same number of realizations and the same random seeds as the simulations ICs. I've tested the impact of the aliasing discussed in \S\ref{sec:ics} by running \texttt{FRMF} with and without zero padding, and found that the difference is negligible with the PNG models used here. Thus, in all the calculations zero padding was not used.

Both \texttt{FRMF} and \texttt{BanyanPM} were run on the same machine, a mini-pc with the AMD Strix Halo architecture and 128 GB of ``unified'' memory. This machine allowed us to utilize GPU acceleration on large enough meshes, up to $N=1024$ for \texttt{FRMF}. The performance of the codes is summarized in Table~\ref{tab:performance}, and it shows a significantly shorter times for \texttt{FRMF} in comparison with simulations.

\begin{table}[]
    \centering
    \caption{Time for a single realization run of simulation codes and \texttt{FRMF}.}
    \begin{tabular}{lc}
    \hline
    Code \& task   & time \\
    \hline
    \texttt{GADGET-2}, $512^3$ particles, $L=100$~Mpc/h simulation & 50~h \\
    \texttt{BanyanPM}, same simulation & 7.5 h \\
    \texttt{FRMF}, make Gaussian ICs, N=512 & 13 s \\
    \texttt{FRMF}, make non-Gaussian ICs, N=512 & 28 s \\
    \texttt{FRMF}, exceedance + constant barrier, 4 redshifts & 9 s \\
    \texttt{FRMF}, shoreline + ODE table, 4 redshifts & 39 s \\
    \texttt{FRMF}, peak-patch + ODE table, per redshift & 18 s \\
    \texttt{PINOCCHIO} & 280 s \\
    \hline
    \end{tabular}
    \label{tab:performance}
\end{table}

\subsection{Mass function comparison}
\label{sec:sim-mf}
\begin{figure}
    \centering
    \includegraphics[width=0.99\linewidth]{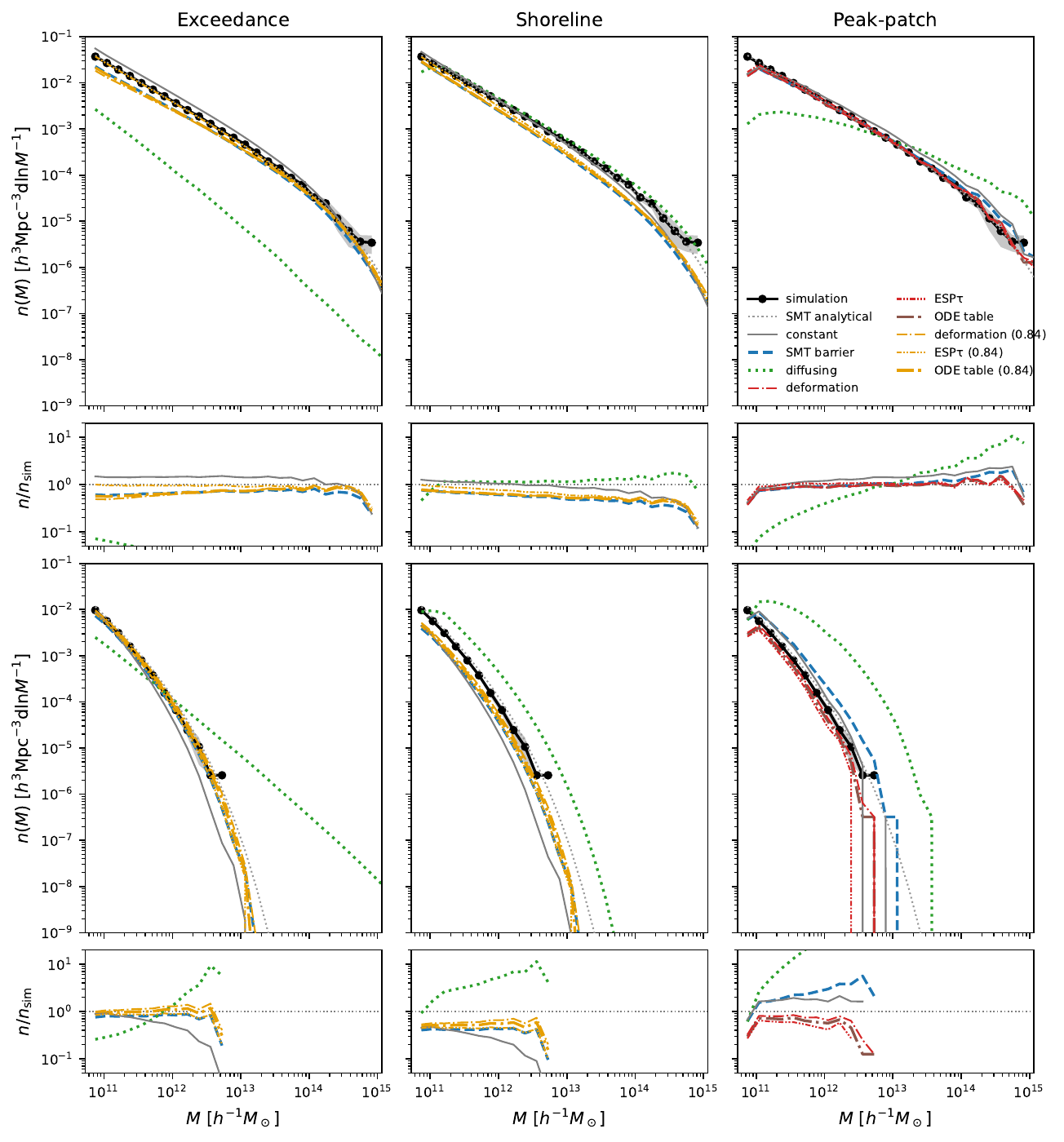}
    \caption{Comparison of simulated mass functions in the Gaussian case in a 100~Mpc/h box with three mass function methods and 6 barriers. As a reference, the SMT \cite{ShethMoTormen2001} analytical formula is added. Top rows: $z=0$, bottom rows: $z=5$. The shaded region shows one standard deviation scatter between realizations.}
    \label{fig:mf_baldi_gaussian}
\end{figure}

\begin{figure}
    \centering
    \includegraphics[width=0.99\linewidth]{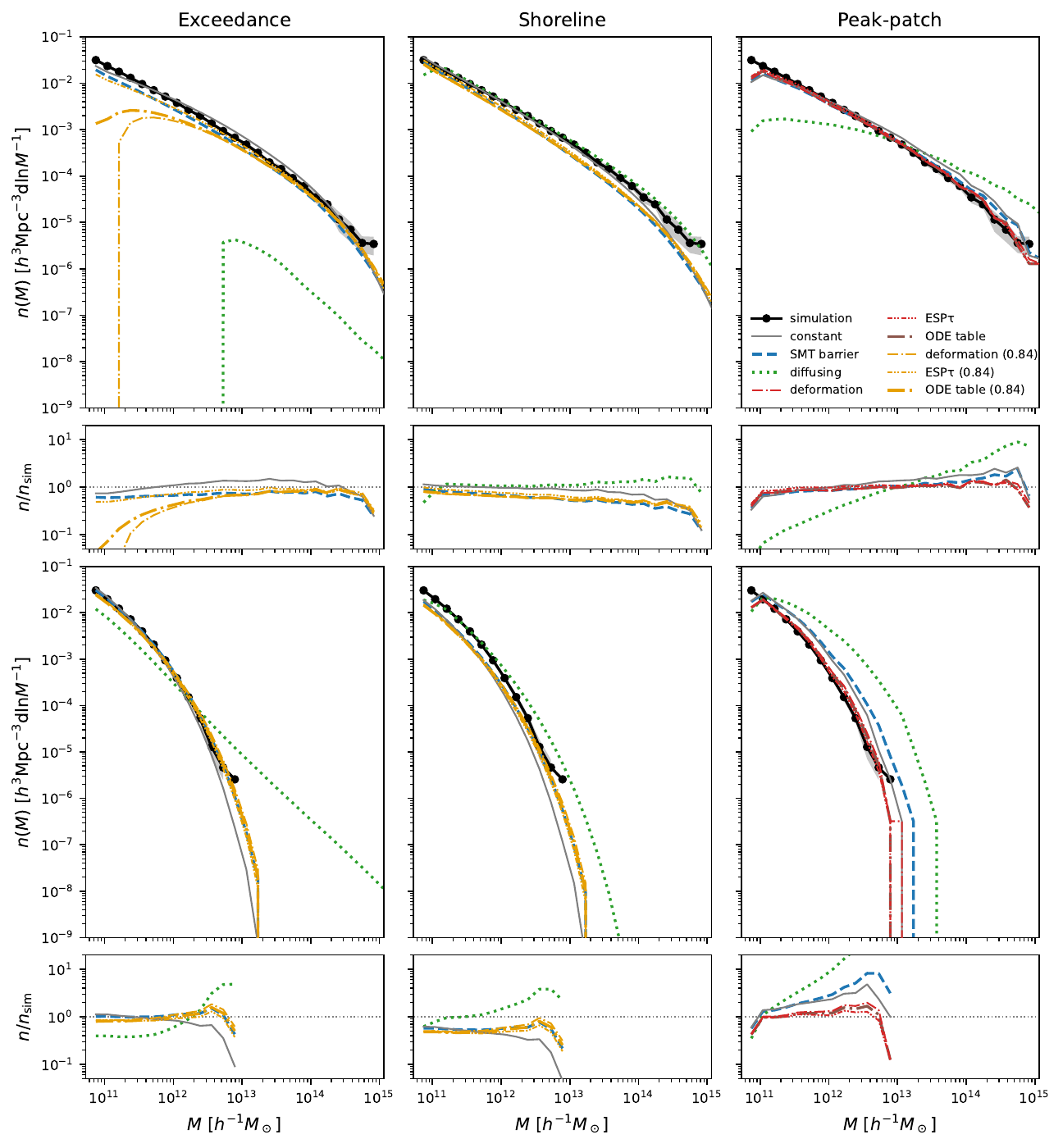}
    \caption{Same as Fig.~\ref{fig:mf_baldi_gaussian}, but for the PNG variant with the $f_{NL}(k)$ shape from Baldi et al \cite{BaldiEtal2024} and a small-scale amplitude of $+10^4$.}
    \label{fig:mf_baldi_p1e4}
\end{figure}

\begin{figure}
    \centering
    \includegraphics[width=0.99\linewidth]{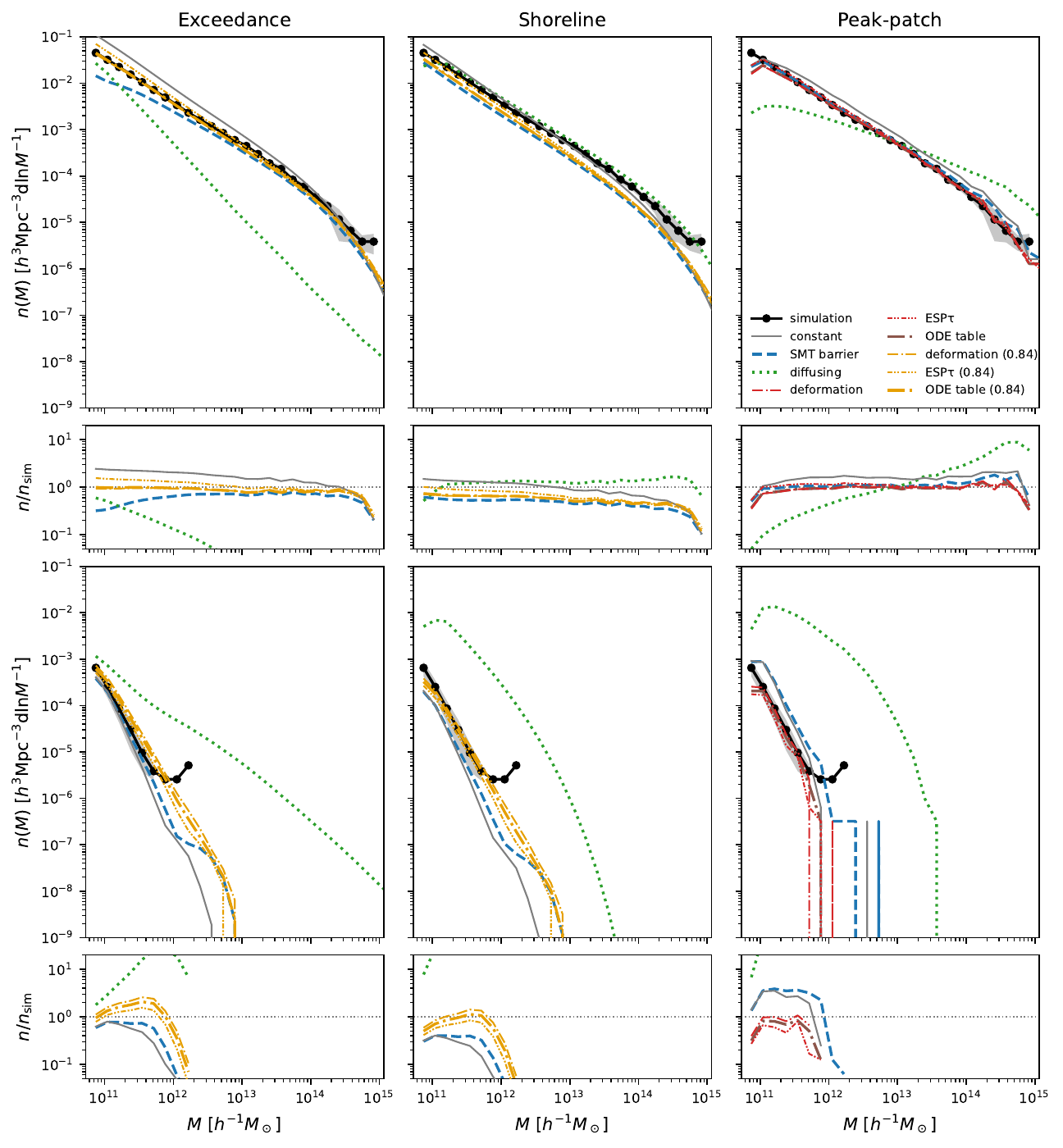}
    \caption{Same as Fig.~\ref{fig:mf_baldi_gaussian}, but for the PNG variant with the $f_{NL}(k)$ shape from Baldi et al \cite{BaldiEtal2024} and a small-scale amplitude of $-10^4$.}
    \label{fig:mf_baldi_m1e4}
\end{figure}

\begin{figure}
    \centering
    \includegraphics[width=0.99\linewidth]{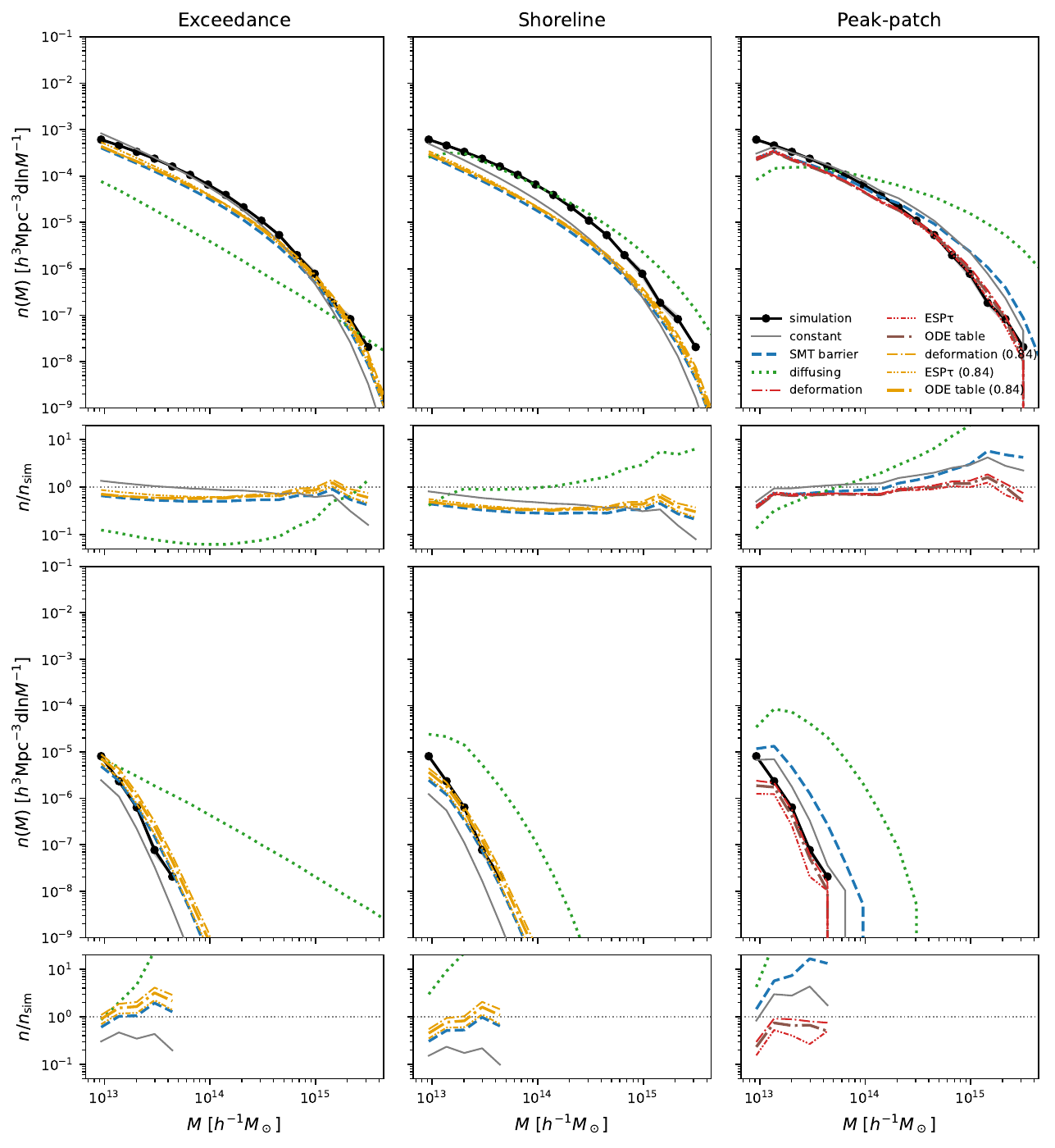}
    \caption{Same as Fig.~\ref{fig:mf_baldi_gaussian}, but for the 500 Mpc/h box and the PNG variant with the $f_{NL}(k)$ shape from Stahl et al \cite{StahlEtal2024b} and a small-scale amplitude of $-600$. The bottom panels are at $z=3$, since at this resolution and at $z=5$ there are too few halos.}
    \label{fig:mf_stahl}
\end{figure}

The resulting mass functions are shown in Figures~\ref{fig:mf_baldi_gaussian}-\ref{fig:mf_stahl}. We focus on the high redshifts, i.e. the bottom two panels on each figure, as this regime is the most challenging and the most sensitive to the barrier choice. From that comparison one can see that the peak-patch method in combination with any of the elliptical barriers provides the most reliable prediction overall of the simulation results. The diffusing barrier works well only for the shoreline method at $z=0$. Surprisingly, the exceedance method which is considered ``incorrect'' (since it has a factor of 2 strictly derived only in the Gaussian case and for a constant barrier), provides a very good prediction for Gaussian and positive $f_{NL}$ PNG cases even at high redshifts, when the SMT or elliptical barriers are used. The negative $f_{NL}$ variants are the most challenging overall, and this is expected: in that case the peak height $\nu$ distribution function is the steepest in the high-$\nu$ regime, and the results are very sensitive to even a slight change in the barrier which cuts high-$\nu$ peaks.

\subsection{Non-Gaussian bias comparison}
\label{sec:sim-bias}
\begin{figure}
    \centering
    \includegraphics[width=0.99\linewidth]{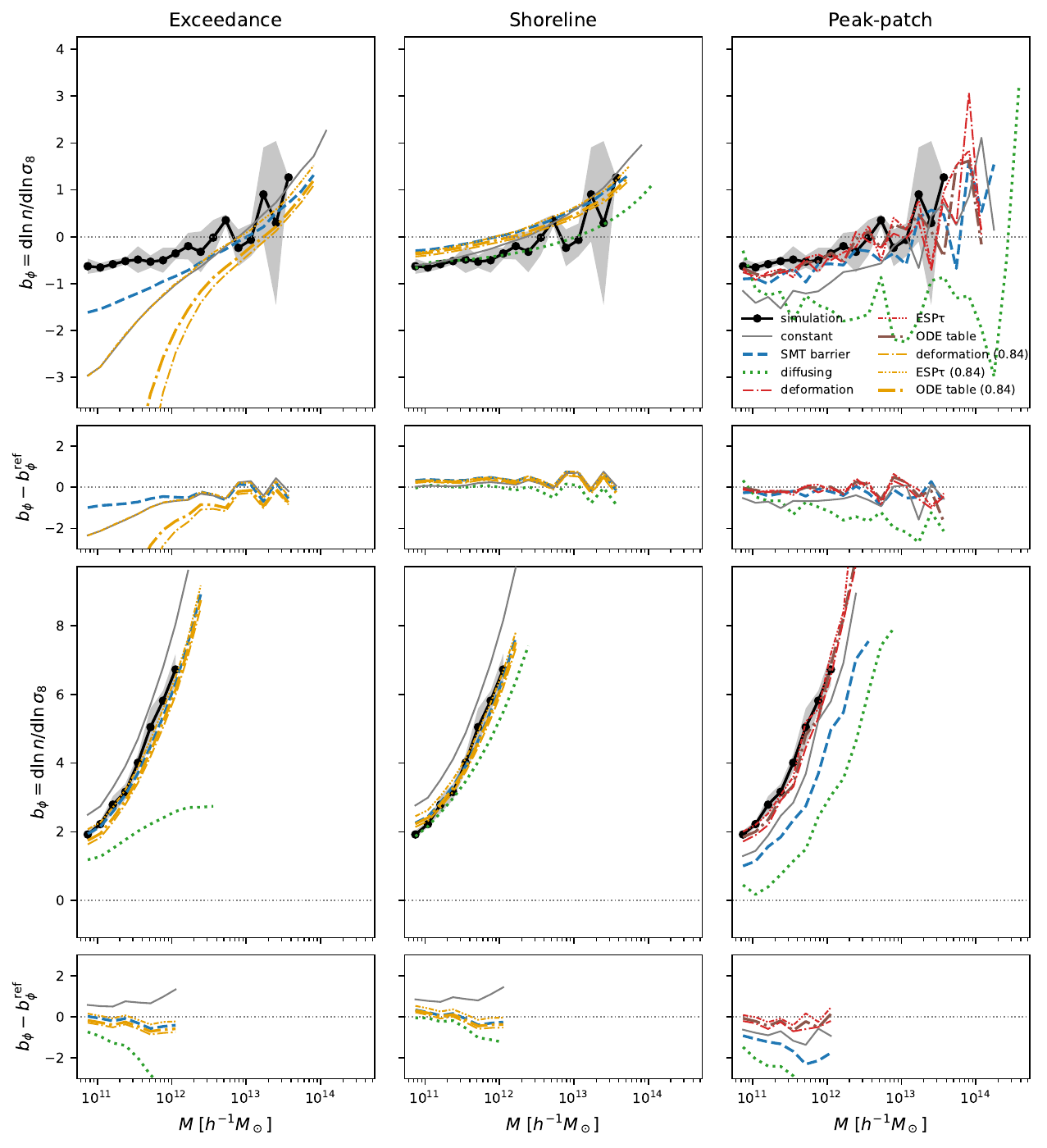}
    \caption{Bias compared with simulations in the Baldi $f_{NL,max}=10^4$ case. Top rows: $z=0$, bottom rows: $z=5$. Shaded region shows one standard deviation scatter.}
    \label{fig:bphi_p1e4}
\end{figure}

\begin{figure}
    \centering
    \includegraphics[width=0.99\linewidth]{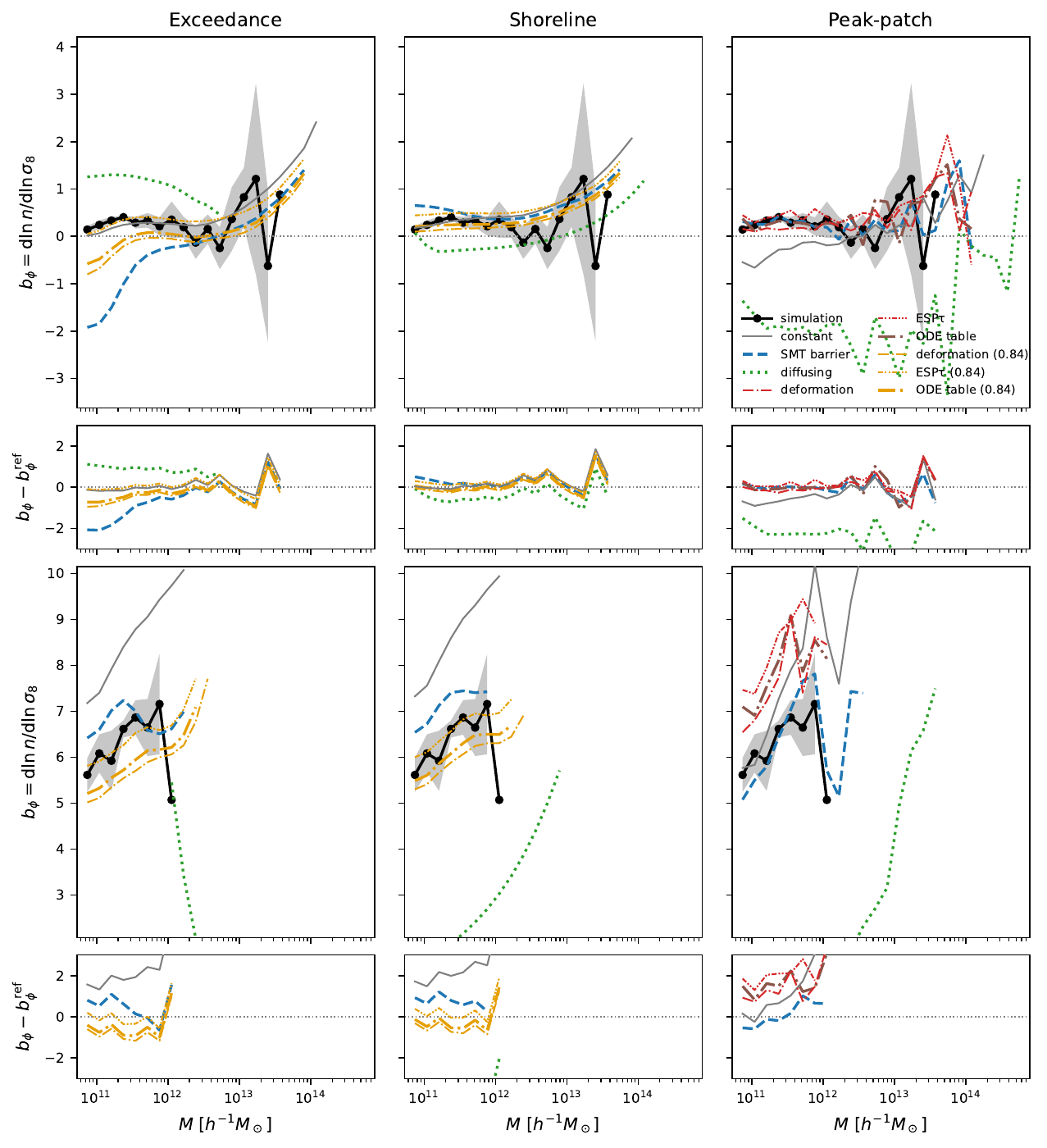}
    \caption{Bias compared with simulations in the Baldi $f_{NL,max}=-10^4$ case. Top rows: $z=0$, bottom rows: $z=3$ (in the negative $f_{NL}$ case there are too few halos at $z=5$ for this investigation). Shaded region shows one standard deviation scatter.}
    \label{fig:bphi_m1e4}
\end{figure}

Besides the MF, we compare the bias parameter measured as 
\begin{equation}
b_\phi = \partial\ln\bar{n}_h/\partial\ln\sigma_8.    
\end{equation}
Numerically we compute the MF in the simulations and FRMF with $\sigma_8 \pm 5$\% to find the derivative. The results are shown in Figs.~\ref{fig:bphi_p1e4}-\ref{fig:bphi_m1e4}. From these figures one can see that the shoreline method with elliptical barriers gives the most consistent match to the simulated bias across two $f_{NL}(k)$ models and two redshifts. As expected from the conclusions of \cite{DesjacquesJeongSchmidt2017}, I found that the elliptical barriers predict $b_\phi$ slightly better than the moving SMT. The diffusing barrier is again worst of all. The exceedance fraction method this time produces a significant offset from the simulations at $z=0$ for any barrier -- this demonstrates that the good match of the MF calculated with this method in Figs.~\ref{fig:mf_baldi_gaussian}-\ref{fig:mf_baldi_m1e4} may be just a coincidence, and changing $\sigma_8$ destroys this match.

The most challenging case for $b_\phi$ prediction is again the large negative $f_{NL}$ model at high $z$ (bottom panels of Fig.~\ref{fig:bphi_m1e4}). The peak-patch method with any of the three elliptical barriers gives a significant offset here, however, the bias itself is high and this offset is only 30\% of the bias value.

\subsection{Adjusting barrier height}
\label{sec:sim-fc}
With N-body simulations we also address the question of the choice of the barrier height factor $f_c$. BM96 \cite{BondMyers1996} propose to set it to 1.0 for their peak-patch method (while elliptical collapse equations give 0.95 in the spherical limit), and SMT \cite{ShethMoTormen2001} propose $f_c=0.84$ which is applicable to our exceedance and shoreline methods. I try to find the best-matching $f_c$ by minimizing the reduced $\chi^2$ of the \texttt{FRMF} fits to the simulation results. For that I use only the ODE Table barrier and two methods: shoreline and peak-patch. 

Two functions are fitted independently: the mass function itself and the MF ratio of the PNG and Gaussian models. This is motivated by the fact that in simulations the MF itself depends on the halo boundary definition used in the halo-finder. The ratio of the mass functions measures just the response of the halo counts to the change of the cosmological model, hence it is less affected by the halo definition.

\begin{figure}
    \centering
    \includegraphics[width=0.99\linewidth]{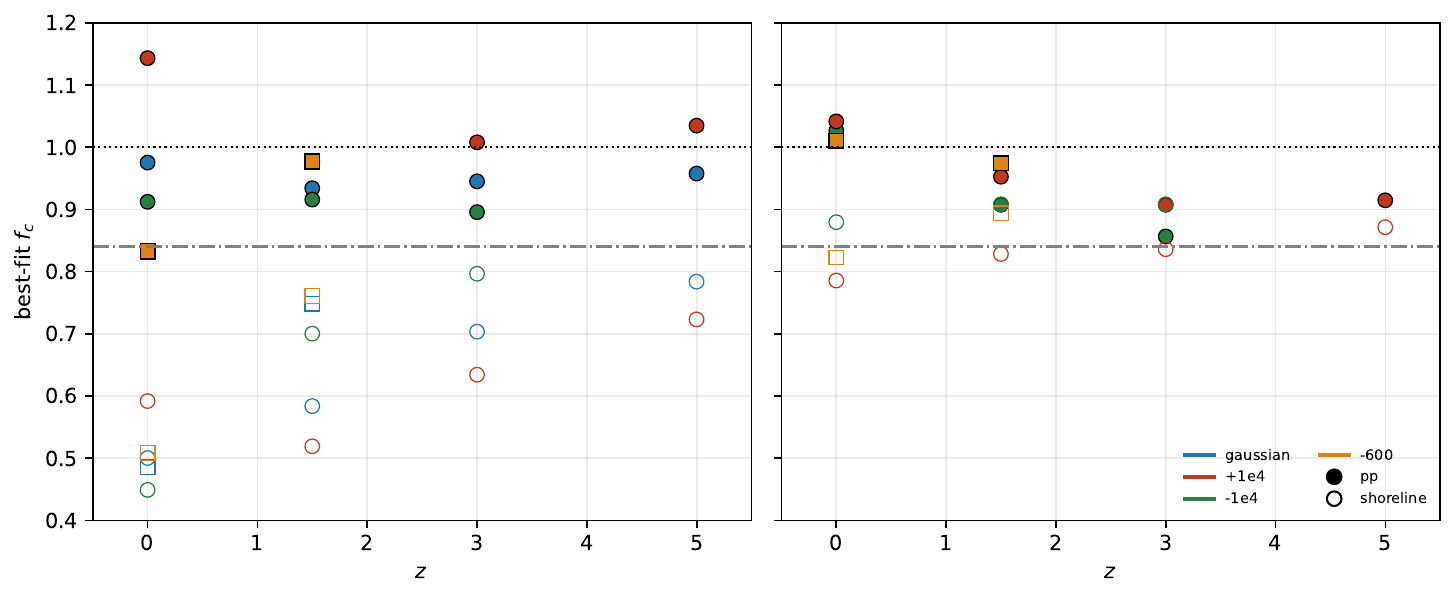}
    \caption{Best-fit barrier height factor $f_c$ for every IC and redshift, for peak-patch (filled symbols) and the shoreline methods (empty symbols). Circles show Baldi-type ICs, squares -- Stahl-type. Left panel -- MF fit, right panel -- MF ratio fit.}
    \label{fig:fc}
\end{figure}

I measure the reduced $\chi^2$ on a grid of $f_c$ from 0.4 to 1.1 in steps of 0.1 and interpolate it with polynomials to find the minimum for each redshift and IC type. The results are shown in Fig.~\ref{fig:fc}. First consider the left panel on which the $f_c$ that fits the simulated MFs for various models and redshifts is shown. From this panel one can see that for the peak-patch method there is no obvious redshift/ICs type trend, and the mean value is actually 0.948, and the median 0.957. The mean value surprisingly agrees well with the raw ODE calculation result, so the re-normalization done in BM96 actually makes the results slightly worse. 

For the shoreline method, the mean is 0.595 and the median is 0.634 -- well below the 0.84 proposed by SMT. Actually, all the minimum $\chi^2_\mathrm{red}$ points lie below the 0.84 value. Also there is quite an obvious trend with redshift, with smaller $f_c$ at low $z$. These findings argue in favor of the peak-patch method which does not have the trend when the MF prediction is considered.

On the right panel of Fig.~\ref{fig:fc} the values of $f_c$ were found as the best fit to the MF ratio for the PNG and Gaussian models. Here the results show the opposite behavior in comparison with the left panel: there is an obvious trend in the $f_c$ for the peak-patch method (but the mean is 0.92, and the median is 0.95), while for the shoreline method the best $f_c$ is constant, with a mean of 0.86 and a median of 0.87. The scatter of the results makes the SMT value of 0.84 actually consistent with these results. So when the goal is to predict the MF ratio, the shoreline method with $f_c=0.84$ is the favorite.

\subsection{Comparison with analytical approximations}
\label{sec:sim-approx}
In \S\ref{sec:sim-mf}-\ref{sec:sim-bias} I have established that the shoreline and peak-patch methods when applied to the elliptical barriers give the most accurate predictions of the MF and the non-Gaussian bias. In \S\ref{sec:sim-fc} I've also found that the barrier height may be slightly adjusted to improve the accuracy even more in case of the peak-patch method, while for the shoreline, $f_c=0.84$ is justified by fitting the simulation results of the MF ratio. Here I do the final quantitative comparison of those methods with the adjusted barrier height, against simulations, and compare them with the analytic expressions and \texttt{PINOCCHIO} code.


\begin{figure}
    \centering
    \includegraphics[width=0.99\linewidth]{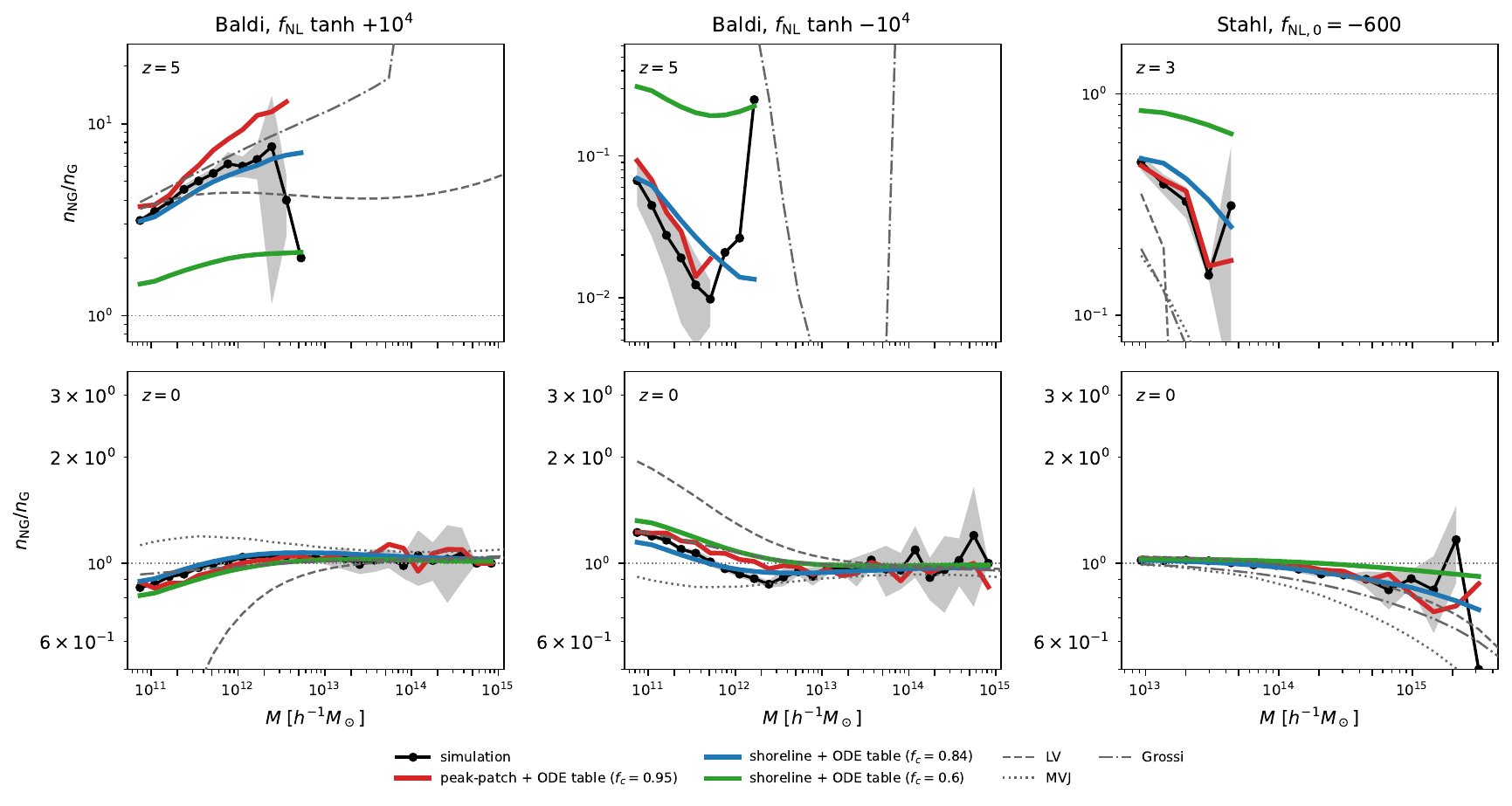}
    \caption{Comparison of the MF ratio of the PNG and Gaussian cosmological models. The black line with points shows the simulation results, with 1 standard deviation scatter shown in grey. Red, blue and green solid lines show the results for \texttt{FRMF} peak-patch + ODE Table ($f_c=0.95$), shoreline + ODE Table ($f_c=0.84$ and $f_c=0.6$), respectively. Thin lines show LV (dashed), MVJ (dotted) and Grossi (dash-dotted) analytical approximations.}
    \label{fig:mfr}
\end{figure}

For the analytical approximations, the comparison here is done in terms of the MF ratio of the PNG and Gaussian cases -- that's what the analytical approximations like MVJ, LV and Grossi introduced in \S\ref{sec:barriers-png} give.
The results for the MF ratio are shown in Fig.~\ref{fig:mfr}. One can see that at $z=0$ in all the considered PNG cases, and for the positive $f_{NL}$ at high redshift the Grossi formula gives an adequate fit over the full available mass range. At high $z$ and for the negative $f_{NL}$ models all analytical approximations fail. From the \texttt{FRMF} side, the ODE Table barrier with the peak-patch method and $f_c=0.95$, as well as the shoreline method with $f_c=0.84$ produces comparable results which closely reproduce the simulation results for all models and redshifts. The case with $f_c=0.6$ which was found by fitting the MFs, poorly reproduces the MF ratio.

\begin{table}[]
    \centering
    \caption{Accuracy of various methods. \textbf{Methods:} E = exceedance, S = shoreline, PP = peak-patch. \textbf{Barriers:} const = constant, ST = moving, diff = diffusing, def = deformation (Sandvik), ESP = ESP$\tau$, ODE = ODE Table. The non-default barrier height adjustment factor $f_c$ is shown in parentheses. For the MF and MF ratio the relative deviation is shown, $\exp(|\log(y/y_{ref})|)$. Med = median, max = maximum over all ``good'' bins, which are defined in the paper text. Fail is the percentage of bins with good simulation results and NaN returned by a method. For the bias parameter, the absolute deviation is shown.}
    \begin{tabular}{lccc}
\hline
method / barrier & MF & MF ratio & $b_\phi$ \\
                & med / max / fail\% & med / max / fail\% & med / max \\
\hline
E / const & 1.32 / 2.36 / 0 & 1.12 / 1.74 / 0 & 0.16 / 4.7 \\
E / ST & 1.40 / 4.23 / 0 & 1.09 / 3.11 / 0 & 0.42 / 2.7 \\
E / diff & 8.01 / 56.3 / 5 & 1.65 / 6.11 / 8 & 1.6 / 30 \\
E / def (0.84) & 1.19 / 9.30 / 0 & 1.14 / 4.89 / 0 & 0.75 / 14 \\
\textbf{E / ESP (0.84)} & 1.20 / 1.77 / 0 & 1.08 / 1.67 / 0 & 0.22 / 2.9 \\
E / ODE (0.84) & 1.22 / 5.05 / 0 & 1.11 / 2.96 / 0 & 0.57 / 27 \\
S / const & 1.55 / 4.35 / 0 & 1.08 / 1.54 / 0 & 0.39 / 4.9 \\
S / ST & 2.24 / 4.91 / 0 & 1.14 / 2.09 / 0 & 0.39 / 2.9 \\
S / diff & 1.33 / 11.3 / 0 & 1.16 / 5.09 / 0 & 0.65 / 9.2 \\
S / def (0.84) & 1.89 / 2.88 / 0 & 1.04 / 1.78 / 0 & 0.28 / 3.2 \\
S / ESP (0.84) & 2.01 / 3.02 / 0 & 1.04 / 1.51 / 0 & 0.40 / 3.0 \\
\textbf{S / ODE (0.84)} & 1.97 / 3.04 / 0 & \textbf{1.04 / 1.68 / 0} & \textbf{0.31 / 2.9} \\
PP / const & 1.61 / 3.16 / 0 & 1.12 / 1.51 / 0 & 0.66 / 3.8 \\
PP / ST & 1.24 / 5.10 / 0 & 1.06 / 1.88 / 0 & 0.57 / 8.8 \\
PP / diff & 3.22 / 69.9 / 0 & 1.30 / 6.26 / 0 & 2.7 / 17 \\
PP / def & 1.13 / 2.92 / 0 & 1.08 / 2.40 / 0 & 0.34 / 3.4 \\
PP / ESP & 1.13 / 4.31 / 0 & 1.11 / 3.06 / 0 & 0.27 / 2.8 \\
PP / ODE & 1.15 / 3.43 / 0 & 1.10 / 2.62 / 0 & 0.32 / 3.1 \\
\textbf{PP / ODE (0.95)} & \textbf{1.11 / 2.43 / 0} & 1.07 / 1.96 / 0 & 0.40 / 4.2 \\
S / ODE (0.60) & 1.38 / 7.40 / 0 & 1.12 / 3.68 / 0 & 0.98 / 12 \\
LV & -- & 1.11 / 5.18 / 0 & -- \\
MVJ & -- & 1.27 / 1.8e3 / 1 & -- \\
Grossi & -- & 1.07 / 1.61 / 7 & -- \\
PINOCCHIO & 1.19 / 3.50 / 0 & 1.35 / 2.82 / 0 & 0.59 / 2.3 \\
\hline
\end{tabular}

    \label{tab:accuracy}
\end{table}

The quantitative comparison of the accuracy of all the methods available in \texttt{FRMF} and for analytical methods and the \texttt{PINOCCHIO} code is shown in Table~\ref{tab:accuracy}. 
This table presents the median and maximal deviations of methods from simulations across all available models and redshifts over the bins which satisfy the following criteria:
\begin{enumerate}
  \item the simulated mass function is finite and positive (the NaN-skipping box mean is $>0$);
  \item at least $100$ simulated halos fall in the bin, summed over all realizations;
  \item at least half the realizations are occupied (a finite positive count);
  \item $M \ge 3\,M_{\rm min}$;
  \item the FRMF prediction has at least $50$ halos.
\end{enumerate}
For the ratio, the criteria 1,2,3,5 are applied to \emph{both} the non-Gaussian and the Gaussian legs (and the first rule is important only for the ratio, for the MF itself it is guaranteed by the second rule). For the analytic approximations (LV, MVJ, Grossi) the last criterion is replaced by a finite-prediction guard. Reported values for the MF and the MF ratio are the symmetric relative deviation: $\exp(|\log(n_\mathrm{method}/n_\mathrm{sim})|)$ (and the analogous quantity for the ratio), pooled over all models, redshifts and bins, summarized as median / maximum / percentage of failed bins. The latter are the bins where the simulations produce results which satisfy the criteria mentioned above, while the method produces NaN or zero. For the bias parameter $b_\phi$ the absolute deviations are reported.

Three methods in Table~\ref{tab:accuracy} are highlighted, they are considered the ``best'' ones for several reasons. The peak-patch method with the adjusted barrier height ($f_c=0.95$) provides the most reliable fit to the simulated MF across all the models and redshifts. This is seen from Figs.~\ref{fig:mf_baldi_gaussian}-\ref{fig:mf_baldi_m1e4} and the median MF deviation in Table~\ref{tab:accuracy}. The shoreline method with the ODE table barrier and the SMT-motivated height $f_c=0.84$ provides the best fit to the MF ratio and at the same time to the bias parameter, so this method best captures the non-Gaussian features. The difference with the Sandvik barrier is small, so those methods can be considered equivalent.

Interestingly, some of the exceedance fraction method variants give an even better median $b_\phi$ deviation, e.g. exceedance + constant barrier. However, looking at Figs.~\ref{fig:bphi_p1e4} and \ref{fig:bphi_m1e4}, left panels, solid gray line, one can see that this method fails badly in some cases, like the Baldi case with $f_{NL}=+10^4$ at $z=0$. ESP$\tau$ barrier with the exceedance fraction method also provides quite good median deviations in Table~\ref{tab:accuracy}, but again fails in the $f_{NL}=+10^4$ at $z=0$ case. The reason why the ``wrong'' exceedance method with a ``manual'' factor of 2 gives very small deviations in some non-Gaussian cosmologies is unclear.

For the analytical methods, the Grossi formula is the most accurate for the cases we consider, but in some cases it fails to give an answer (there are 7\% of failed bins in total). This is demonstrated in Fig.~\ref{fig:mfr}, top row, second column, where the Baldi case with negative $f_{NL}$ is presented, and the Grossi formula cannot be computed in the simulated range.

The \texttt{PINOCCHIO} code gives results almost on par with the best \texttt{FRMF} methods for the mass function and $b_\phi$, but is significantly worse for the MF ratio. The more detailed analysis shows that in \texttt{PINOCCHIO} all the high-z MFs (both Gaussian and non-Gaussian) deviate significantly more than those of \texttt{FRMF}, while at $z=0$ the results are almost identical, and better than \texttt{FRMF} in the Gaussian case.

\section{Conclusions}
\label{sec:conclusion}
The code presented in this paper, \texttt{FRMF}, provides machinery to compare different methods of halo mass function computation using the same realizations of random Gaussian and non-Gaussian density fields. At the time of writing the paper, the code contains three MF calculation methods: the exceedance fraction, the shoreline and the peak-patch, as well as the output of initial conditions for running cosmological simulations. The three methods can be computed much faster than simulations: there are no time steps or particle-trajectory integration, the density field is just smoothed with the top-hat filter and some barrier is applied to the smoothed field. Each of the three methods can be used with one of the six available barriers: from the simple Press-Schechter constant barrier to the elliptical barriers which apply to the combination of the density field and shear tensor eigenvalues.

This paper presents the results of the comparison of the three MF calculation methods and the 6 barriers with simulations in \S\ref{sec:comparison}. The test suite contains realizations of Gaussian initial conditions for two box sizes, and three variants of local non-Gaussianity with scale-dependent $f_{NL}(k)$. The latter were chosen to reach high values, $\sim 10^3-10^4$ at small scales not constrained by CMB maps. From this comparison, several observations are made:
\begin{itemize}
    \item The best accuracy of the prediction of the mass function, the non-Gaussian bias and the mass function ratio for PNG and Gaussian models is reached with the elliptical barriers with either the shoreline or the peak-patch method.
    \item The peak-patch method with the ODE table barrier and $f_c=0.95$ provides predictions of the mass function with a median deviation of 11\% across all the realizations, redshifts and PNG models.
    \item The shoreline method with the ODE table barrier and $f_c=0.84$ provides a good fit to the MF ratio, with a median deviation of only 4\% and a maximum deviation up to 60\%. At the same time, this method also better predicts the bias parameter $b_\phi$ than the peak-patch method.
    \item The $f_c=0.84$ value for the shoreline method with the ODE table barrier is taken from the SMT argument, and is also confirmed by fitting the MF ratio to simulations in this paper. 
    \item The exceedance fraction method provides a good fit to the MF, but only in some combinations of the barrier and the PNG type. E.g. the ODE Table barrier is a good fit for Baldi-type PNG with negative $f_{NL}$, but a poor fit to the positive $f_{NL}$ model at $z=0$.
\end{itemize}

My final recommended recipe to calculate the MF in non-Gaussian cosmological models with a large $f_{NL}$ at small scales is the following. Use the Gaussian MF from a simulation or a well-tested fit as a reference. Using \texttt{FRMF} calculate the shoreline + ODE table (with $f_c=0.84$) MF ratio and multiply it by the reference Gaussian MF. This guarantees the MF difference from real non-Gaussian ICs simulations of the order of few \% at $z=0$ and less than few tens \% at high redshifts.

\section*{Acknowledgements}
This work is supported by the RSF grant 25-22-00763 ``Simulations of populations of cosmological voids'' \url{https://rscf.ru/en/project/25-22-00763/}. 

This work made extensive use of the Julia programming language \cite{bezanson2017julia} together with the open-source packages FFTW \cite{frigo2005fftw} for fast Fourier transforms, 
\texttt{OrdinaryDiffEq.jl} \cite{rackauckas2017differentialequations} for the ellipsoidal-collapse ODE integration, and HDF5 \cite{hdf5} for data storage. Additional Julia packages used include QuadGK, LoopVectorization, SpecialFunctions, ProgressMeter, PythonPlot and IJulia.

\section*{Declaration of generative AI and AI-assisted technologies in the manuscript preparation process}

During the preparation of this work the author used the AI models Claude Sonnet 4.6 and Claude Opus 5 (via \texttt{Claude Code}) and DeepSeek V4 Pro (via \texttt{OpenCode}) to write a substantial part of the code, manage the running of simulations, and gather related literature (using the LLM Wiki pattern). After using these tools, the author reviewed and edited the content as needed and takes full responsibility for the content of the published article.

\def\apj{Astrophys.~J}
\def\apjl{Astrophys.~J.,~Lett}
\def\apjs{Astrophys.~J.,~Supplement}
\def\an{Astron.~Nachr}
\def\aap{Astron.~Astrophys}
\def\mnras{Mon.~Not.~R.~Astron.~Soc}
\def\pasp{Publ.~Astron.~Soc.~Pac}
\def\aaps{Astron.~and Astrophys.,~Suppl.~Ser}
\def\apss{Astrophys.~Space.~Sci}
\def\ibvs{Inf.~Bull.~Variable~Stars}
\def\japa{J.~Astrophys.~Astron}
\def\na{New~Astron}
\def\aspproc{Proc.~ASP~conf.~ser.}
\def\aspcs{ASP~Conf.~Ser}
\def\aj{Astron.~J}
\def\actaa{Acta Astron}
\def\araa{Ann.~Rev.~Astron.~Astrophys}
\def\caosp{Contrib.~Astron.~Obs.~Skalnat{\'e}~Pleso}
\def\pasj{Publ.~Astron.~Soc.~Jpn}
\def\memsai{Mem.~Soc.~Astron.~Ital}
\def\astl{Astron.~Letters}
\def\aipproc{Proc.~AIP~conf.~ser.}
\def\physrep{Physics Reports}
\def\prd{Phys. Rev. D}
\def\jcap{Journal of Cosmology and Astroparticle Physics}

\bibliographystyle{elsarticle-num}
\bibliography{references}
\end{document}